\documentclass[11pt]{article}
\usepackage{amssymb}

\newcommand{\C}{\mathbb{C}}

\newcommand{\beq}{\begin{equation}}
\newcommand{\eeq}{\end{equation}}
\newcommand{\beqn}{\begin{eqnarray}}
\newcommand{\eeqn}{\end{eqnarray}}

\newcommand{\set}[1]{{\left\{{#1}\right\}}}

\newcommand{\la}{\lambda}
\newcommand{\si}{\sigma}

\newcommand{\eps}{\epsilon}

\newcommand{\be}{\beta}

\newcommand{\te}{\theta}

\newcommand{\pa}{\partial}
\newcommand{\al}{\alpha}

\newcommand{\del}{\delta}
\newcommand{\lga}{\longrightarrow}

\newcommand{\da}{\dagger}

\newcommand{\ov}{\over}

\newcommand{\lb}{\label}

\newcommand{\NP}[1]{ {\it Nucl.~Phys.} {\bf #1}}

\newcommand{\PR}[1]{ {\it Phys.~Rev.} {\bf #1}}
\newcommand{\PRL}[1]{ {\it Phys.~Rev.~Lett.} {\bf #1}}

\newcommand{\JP}[1]{ {\it J.~Phys.} {\bf #1}:\  Math.~Gen.~}

\begin{document}
\begin{titlepage}
\setcounter{page}{1}
\renewcommand{\thefootnote}{\fnsymbol{footnote}}

\begin{flushright}
UCDTPG 07-01\\
\end{flushright}

\vspace{6mm}
\begin{center}

{\Large\bf Anomalous Quantum Hall Effect on Sphere}

\vspace{6mm}

{\bf Ahmed Jellal\footnote{ajellal@ictp.it, jellal@ucd.ac.ma}}

\vspace{4mm}

{\em The Abdus Salam International Centre for Theoretical Physics},\\
{\em  Strada Costiera 11, 34014 Trieste, Italy}\\

{\em and} \\

{ \em Theoretical Physics Group,  
Faculty of Sciences, Chouaib Doukkali University},\\
{\em Ibn Ma\^achou Road, P.O. Box 20, 24000 El Jadida,
Morocco}\\

\begin{abstract}

We study the anomalous quantum Hall effect exhibited by 
the relativistic particles 
living on two-sphere $\mathbb{S}^2$ and submitted to 
a magnetic monopole. We start by establishing a direct 
connection between the Dirac and 
Landau operators through the Pauli--Schr\"odinger
Hamiltonian $H_{\sf s}^{\sf SP}$. 
This will be helpful in the sense that
the Dirac eigenvalues and eigenfunctions will be easily derived.
In analyzing $H_{\sf s}^{\sf SP}$ spectrum,
we show that 
there is a composite fermion nature supported by the presence of
two effective magnetic fields.
For the lowest Landau level, we argue that the basic physics of 
graphene is similar to that of two-dimensional electron gas,
which is in agreement with the planar limit.
 For the higher Landau levels,
we propose a $SU(N)$ wavefunction for different filling factors
that captures all symmetries. Focusing on the graphene case, i.e. $N=4$, 
we give different configurations
those allowed to recover some known results.

\end{abstract}

\end{center}
\end{titlepage}



\section{Introduction}

Since its discovery, the quantum Hall effect (QHE)~\cite{prange} is always 
bringing new surprises. In fact,
recently with the new technological progress, one can observe
the effect in graphene~\cite{novoselov,zhang}, which is
a two-dimensional (2D) projection of graphite. It is governed
by a Hall conductivity of the form  
$$\si_{H}^{\sf g}=4\left(l+{1\over 2}\right){e^2\over h}, \qquad 
l =0, \pm1, \pm 2\cdots$$
which was predicted theoretically~\cite{ando,sharapov}. It is
different from that characterizing the integer quantum Hall effect
(IQHE) exhibited by  two-dimensional electron
gas (2DEG) in the presence of an external magnetic field~\cite{klitzing}:
$\si_{H}^{\sf sc}= l {e^2\over h}$ where $l\in \mathbb{N}$.
The anomalous Hall conductivity 
$\si_{H}^{\sf g}$
can be seen as 
a consequence of different effects. 
Indeed,
the prefactor $4$ reflects the two-fold spin and
two-fold valley degeneracy in the graphene band structure. However, 
the term ${1\over 2}$ comes from the Berry phase due to the pseudospin 
(or valley) precession
when a massless (chiral) Dirac particle exercises cyclotron motion. 
Moreover, $\si_{H}^{\sf g}$ 
provides a direct evidence for the relativistic
nature of the charge carriers in graphene~\cite{yang}. 
 This offers unexpected bridge
between condensed matter physics and quantum electrodynamics,
more detail can be found in~\cite{katsnelson}.

The above new challenge offered a laboratory for different
investigations. Theoretically, many works have been 
reported on the subject and extended to the fractional
quantum Hall effect (FQHE)~\cite{Tsui82}. 
In fact, a set of
integer and FQHE states that break the $SU(4)$ spin/valley symmetry
has been constructed~\cite{yang2}.
Also it is shown that the lowest Landau level (LLL) 
FQHE in graphene, in the large Zeeman energy limit, is equivalent to the LLL
in 2DEG~\cite{toke}. Using the composite fermion (CF)
theory~\cite{cf}, different developments have been appeared dealing with
a possible FQHE in graphene. More precisely, a direct and immediate
mapping between IQHE resulting from CF of graphene and FQHE has been
established~\cite{castroneto}. New FQHE states that have no analogue in 2DEG
have been predicted~\cite{toke2} and also
 compressible states at $\nu=\pm{1\over 2}$ and
 $\nu=\pm{3\over 2}$ have been discussed~\cite{khvesh}. Furthermore,
Using the Halperin theory~\cite{halperin} for spin in 
the conventional QHE, 
a $SU(N)$ wavefunction has
been constructed and related discussions have been 
reported by Goerbig and Regnault~\cite{goerbig2}. Other 
important investigations related to 
the subject can be found in~\cite{COgraphene,AC}
as well as~\cite{cliff}. For early works, one may see
the papers cited in~\cite{vozmediano}.

Motivated by different analysis of the anomalous fractional 
quantum Hall effect (AFQHE) on plane, we discuss its realization
on two-sphere $\mathbb{S}^2$.
As far as we know, 
such consideration can be argued by the symmetry conservation.
Recalling that 
the Laughlin theory~\cite{laughlin} is translationally invariant but 
not rotationally. To solve such  problem,  Haldane~\cite{haldane} 
analyzed the Hall system on $\mathbb{S}^2$ and constructed a
theory that possess all symmetries. 
This leads to conclude that the study 
of AFQHE on the compact surface will be interesting. Certainly this  
will bring new ideas and 
 show the difference as well as similarities with respect to the standard case.

To do our task, we consider a mathematical formalism governed by
the Pauli--Schr\"odinger and
Dirac Hamiltonian's. The spectral properties of the first one
have been considered at many occasions where the asymptotic behavior
of the magnetic field at infinity has been discussed~\cite{shigekawa}. Also
the eigenprojector kernels have been analytically studied~\cite{ghanmi}. 
The second one has been analyzed 
in different contexts, e.g.~\cite{camporesi,bar}. These
 investigations show that both Hamiltonian's are 
interesting from mathematical point view. Their extensions to
physics have been done for different motivations,
e.g. the anomalous QHE on plane. In the present
paper, we give another way to show the physical realization
of both Hamiltonian's. This concern a survey of
the anomalous QHE on $\mathbb{S}^2$.

It is well-known that the Landau problem is the cornstone of the conventional QHE.
However, for the anomalous QHE, one could start with the Dirac Hamiltonian because
particles are relativistic. Both of systems are connected, which
is due to the fact that the anomalous Hall conductivity can be understood from the 
Landau level structures of Dirac particles. Because of this connection, we start
by considering the Pauli--Schr\"odinger Hamiltonian
describing particles on $\mathbb{S}^2$ in the presence of
a magnetic monopole. 
Its matrix elements are obtained to be dependents of two different effective
magnetic fields. This will offer another way to make contact with
the composite fermions. From this analysis,
we derive the Dirac eigenvalues as well as
the corresponding eigenfunctions.

We give some discussions about the Pauli--Schr\"odinger spectrum. Indeed,
in determining such spectrum we show that
its matrix elements are expressed in terms of the Landau Hamiltonian.
Each element describes a subsystem subjected to an effective magnetic field.
The sum of both parts gives exactly the external magnetic field. We argue
that these fields are similar to those felt by the composite fermions
in 2DEG~\cite{cf}. This interpretation allows us to define an effective filling factor
and establish different links with
other theories.
 
We start our analysis of the anomalous QHE by considering LLL. Indeed,
to discuss similarities and differences
between FQHE on 2DEG and graphene, we evaluate some physical quantities.
In doing these,
we show that AFQHE in both systems is similar
when the 
particle are restricted
to live on LLL. To argue this, we
evaluate the incompressibility through the correlation
function and determine the density of particles.
As interesting results, we conclude that the basic physics in
the graphene LLL is the same as in 2DEG LLL for the Hall
system on $\mathbb{S}^2$. This coincides with the planar limit
analysis~\cite{toke}.

 As far as the higher Landau levels are concerned, 
Goerbig and Regnault~\cite{goerbig2} proposed a
wavefunction to describe $SU(4)$ FQHE in graphene sheet. This
is seen as a direct extension of the Halperin wavefunction~\cite{halperin}.
We note that the proposed state is related to our ground state
configuration~\cite{schreiber} realized in terms of the matrix model
and non-commutative Chern--Simons theories. To consider
the higher Landau levels on $\mathbb{S}^2$, we
 build a $SU(N)$ wavefunction
generalizing those of Haldane~\cite{haldane} and
 corresponding to different filling factors~\cite{wen}
$$\nu = q_iK_{ij}^{-1}q_j^{-1}$$
where $K_{ij}$ is $N\times N$ matrix and $q_i$ is a vector.
To make contact with graphene, we analyze the $N=4$ case
and show its relations to others theories. For this,
we make different appropriate choices of matrices
to reproduce specific filling factors.

The paper is organized as follows. In section 2, we discuss the
anomalous IQHE on the plane, which can be done by introducing
 the corresponding Hamiltonian formalisms. In section 3, 
we introduce the covariant derivatives acting on $p$-forms of
$\mathbb{S}^2$ to write algebraically and analytically the
Pauli--Schr\"odinger Hamiltonian $H_{\sf s}^{\sf PS}$. 
Using an unitary transformation,
we show that  $H_{\sf s}^{\sf PS}$ can be
mapped in terms of that of Landau. This mapping will be used
to derive the Dirac spectrum. In section 4, we show that 
the Pauli--Schr\"odinger system is exhibiting a composite fermion
behavior where the effective magnetic fields will be defined.
Subsequently, we discuss
AFQHE on $\mathbb{S}^2$ in terms of the density of particles and two-point
function. 
We generalize the Haldane wavefunction and reproduce different
filling factors in section 5. 
We conclude and give some perspectives
in the final section.

\section{Anomalous QHE on plane}

Before developing our main ideas, it is relevant
to discuss
the anomalous IQHE on the plane and emphasis its differences
with respect to the conventional QHE. 
Since this new challenge is due to a manifestation of the
relativistic particles, it is natural to introduce
the Dirac formalism in $2D$. This can be done
by considering the Pauli--Schr\"odinger Hamiltonian and 
resorting its eigenvalues and eigenfunctions in terms of those 
of Landau.

\subsection{Pauli--Schr\"odinger Hamiltonian}

Let us consider one-relativistic particle living
on the plane $(x, y)$  in presence of a
perpendicular magnetic field $B$.  
The Pauli--Schr\"odinger Hamiltonian for such system
can be written as
\beq
H_{\sf p}^{\sf PS}= {1\over 2m} \left[\vec\si\cdot\left(\vec p - 
{e\over c} \vec A\right)\right]^2
\eeq
where the Pauli matrices $\vec\si$ satisfy the usual relations
\beq
\left\{\si_i, \si_j\right\} = 2\del_{ij}, \qquad 
\left[\si_i, \si_i\right] = 2\eps_{ijk} \si_k.
\eeq
It will be clear that
the square root of $H_{\sf p}^{\sf PS}$ gives exactly
the Dirac Hamiltonian in $2D$. This connection will 
be used in the next in order to achieve our purpose.

For simplicity, it is convenient to express
$H_{\sf p}^{\sf PS}$  in  terms of
the Landau Hamiltonian $H^{\sf L}_{\sf p}$. Indeed, by choosing
the symmetric gauge
\beq\lb{syga}
\vec A={B\over 2}\left(-y, x\right)
\eeq
we show that $H_{\sf p}^{\sf PS}$ can be mapped as
 \beq 
H_{\sf p}^{\sf PS}  =
\left(\begin{array}{cc}
H^{\sf L}_{\sf p} & 0\\ 0 & H^{\sf L}_{\sf p}\end{array} \right) 
-{B}
\left(\begin{array}{cc}1 & 0\\ 0 & -1
\end{array}\right)
\label{PHC}
\eeq
where $H^{\sf L}_{\sf p}$, in complex notation $z=x+iy$,
takes the form
\beq
H^{\sf L}_{\sf p}= -\left\{\frac{\partial^2}{\partial z
\partial \bar z}+{B}  \left(z\frac{\partial}{\partial z}-
\bar z \frac{\partial}{\partial \bar z}\right)-{B}^2 z\cdot \bar z\right\}.
\label{LHC}
\eeq
Hereafter, we set the fundamental constants $(e, c,\hbar, m)$ to one.
Clearly, the spectrum and eigenfunctions of
 $ H_{\sf p}^{\sf PS} $
can be derived from that of $H_{\sf p}^{\sf L}$. This will play an important
role when we consider the present system on $\mathbb{S}^2$. 

One can also use the algebraic approach to write another version of
$ H_{\sf p}^{\sf PS}$. This can be done by introducing the second order
differential operators 
\beq\lb{pcder}
D_{\sf p} ={\pa} +i{B\over 2} \ \mbox{ext}(\theta_{\sf p}),\qquad
 D_{\sf p}^* ={\pa}^*-i{B\over 2}\ \mbox{ext}(\theta_{\sf p})^*
\eeq
in terms of the gauge potential 
\beq
\theta_{\sf p} = i(\bar z dz - z d\bar z).
\eeq
Recalling that, the wavefunctions of $ H_{\sf p}^{\sf PS} $ 
are type differential one-forms of $\C$ and each one has
two component spinors. They are 
\beq
\Psi = 
\left(\begin{array}{c}
 \Psi_1 \\ \Psi_2
\end{array} \right).
\eeq
The term $\mbox{ext}(\theta_{\sf p})$ acts on the states
as follows
\beq
\mbox{ext}(\theta_{\sf p})\ \Psi =
\theta_{\sf p} \wedge \Psi.
\eeq
Therefore, we can write $ H_{\sf p}^{\sf PS} $ 
in terms of the covariant derivatives as
\beq 
 H_{\sf p}^{\sf PS} = D_{\sf p}^*D_{\sf p} + D_{\sf p}D_{\sf p}^*.
\label{PHRealization}
\eeq 
With this we now have two possibilities to get the eigenvalues
and eigenfunctions of 
$ H_{\sf p}^{\sf PS}$. This can be done either algebraically or analytically.

\subsection{Dirac Hamiltonian}

The above tools can be applied to analyze the anomalous
QHE in graphene. Indeed in such systems, 
the two Fermi points, each with a two-fold band
degeneracy, can be described by a low-energy continuum approximation
with a four-component envelope wavefunction whose components are
labeled by a Fermi-point pseudospin $= \pm 1$ and a
sublattice forming an honeycomb. Specifically, 
the Hamiltonian for one-pseudospin component
can be obtained from (\ref{PHC}) under some considerations. This  
 is~\cite{shon-ando,Semenoff}
\begin{equation}
H^{\sf D}_{\sf p}=v_{\sf F}
\left(\begin{array}{cc}
0 & \Pi_x-i\Pi_y \\
\Pi_x+i\Pi_y & 0 \\
\end{array}\right)
\end{equation}
where $v_{\sf F}\approx {c\over 100}$ is the Fermi velocity and the
many-body effects are neglected. The
conjugate momentum, in the symmetric gauge, are given by  
\beq
\Pi_x=p_x-{B\over 2}y, \qquad \Pi_y=p_y + {B\over 2}x.
\eeq

Using the $H^{\sf D}_{\sf p}$ form, we can establish a link to 
$H^{\sf PS}_{\sf p}$. To proceed,
let us return to (\ref{pcder})
and write explicitly the raising and lowering operators as
\beq
D^*_{\sf p}=-2\partial_z + {B\over 2} \bar z, 
\qquad
D_{\sf p}= 2\partial_{\bar z} + {B \over 2}z
\eeq
which satisfy the commutation relation
\beq
[D_{\sf p}, D_{\sf p}^*] = 2B.
\eeq
These can be used to write $H^{\sf D}_{\sf p}$ as
\begin{equation}
H^{\sf D}_{\sf p}=i{ v_{\sf F}}
\left(\begin{array}{cc}
0 & D_{\sf p}\\
 D^*_{\sf p}& 0
\end{array}\right).
\end{equation}
Its spectrum can be determined in a simple way
if we introduce its square. This is
\begin{equation}\lb{sdham}
\left(H^{\sf D}_{\sf p}\right)^2= { v_{\sf F}^2}
\left(\begin{array}{cc}
 D_{\sf p} D^*_{\sf p}& 0\\
0 & D^*_{\sf p}D_{\sf p}
\end{array}\right).
\end{equation}
Hereafter we set $v_{\sf F}$ to one. $\left(H^{\sf D}_{\sf p}\right)^2$
 is related to the Pauli--Schr\"odinger Hamiltonian (\ref{PHC})
up to some multiplicative constants. 
It is clear that, $\left(H^{\sf D}_{\sf p}\right)^2$ is written in 
terms of the diagonal form of the Landau Hamiltonian~(\ref{LHC}).
Therefore, it spectrum
can easily be obtained.

Before deriving the Dirac eigenvalues and eigenfunctions, we make a general statement.
Indeed,
assuming that 
the eigenvalue equation
\beq\lb{oeveq}
\hat O^2 \psi =  \la^2 \psi
\eeq
is satisfied for a given operator $\hat O$ and a vector $\psi$.
Thus, one can simply check  that the vector 
\beq
\phi^{\pm} = \pm \la \psi + \hat O \psi
\eeq
is an eigenvector of $\hat O$ and verify
\beq
\hat O \phi^{\pm} = \pm \la \phi^{\pm}.
\eeq
Therefore, we underline that for $\phi^{\pm}\neq 0$, 
the eigenvalue of $\hat O$ is given by
$\pm \la$.
This can be applied to the Dirac Hamiltonian
and its square. For this,
we start by solving the equation
\beq
\left(H^{\sf D}_{\sf p}\right)^2 \Psi= E \Psi.
\eeq
Since that $\left(H^{\sf D}_{\sf p}\right)^2$ (\ref{sdham}) has to do with the
Landau Hamiltonian~(\ref{LHC}), the wavefunctions $\Psi$ should be written
in an appropriate form. They are
\beq
\Psi_{m,n}=
\left(\begin{array}{c}
\psi_{m-1,n} \\
  \psi_{m,n}
\end{array}\right)
\eeq
where the eigenfunctions $\psi_{m,n}$ 
\begin{equation}\lb{eflps}
\psi_{m,n}(z)=\frac{(-1)^m\sqrt{B^mm!}}{\sqrt{2^{n+1}\pi (m+n)!}}z^n 
L_m^n\left(\frac{z\cdot \bar z}{2}\right) e^{-{B\over 4}z\cdot \bar z}, 
\qquad m, n =0,1,2\cdots
\end{equation}
are describing particles living on $\mathbb{R}^2$ and subjected to the
magnetic field $B$. 
The corresponding Landau levels are given by
\beq\lb{llevels}
(E_{\sf p}^{\sf D})^2_m ={B } \left (2m+ 1\right).
\eeq
Using the general argument stated before, we can show that
the normalized eigenfunctions of $H^{\sf D}_{\sf p}$
take the form 
\begin{equation}\lb{dwafu}
\Psi_{m\neq0,n}=\frac{1}{\sqrt 2}
\left(\begin{array}{c}
-{\rm sgn}(m)i\psi_{|m|-1,n} \\
\psi_{|m|,n}
\end{array}\right)
\end{equation}
with the convention $\textrm{sgn}(0) = 0$. Note that, the zero-mode 
wavefunction is
\begin{equation}\lb{dwafu0}
\Psi^{(0,n)}=
\left(\begin{array}{c}
0 \\ \psi_{0,n}
\end{array}\right).
\end{equation}
Their energy levels read as
\begin{equation}\lb{dspec}
(E_{\sf p}^{\sf D})_m={\rm sgn}(m)\sqrt{{2  B|m|}}.
\end{equation}
Compared to the usual Landau eigenvalues (\ref{llevels}),
we firstly note that there is a missing factor, i.e. ${1\over 2}$. This is due to
the chirality, which in our case is related to the magnetic field
and is contributing by an amount of $\pm{1\over 2}$.
Secondly unlike the 2DEG LLL, its graphene analogue
is characterized by a zero-energy
and a state given in~(\ref{dwafu0}).

\subsection{Anomalous FQHE on plane}

The Dirac formalism can be employed to show that 
the Hall conductivity on graphene
is different from that on 2DEG. In fact,
for non-interacting electrons, when both the spin and the pseudospin
degeneracies are present, the Hall plateaus are 
\beq\lb{ahcon}
\si_{ H} = 4 \left(l + {1\over 2}\right) {e^2\over 2\pi}, 
\qquad l= 0, \pm 1, \pm 2, \cdots. 
\eeq
This result has been theoretically predicted~\cite{ando,sharapov}
and experimentally seen~\cite{novoselov,zhang}. It can be also obtained by using
the thermodynamical analysis of the Hall particles on 
graphene~\cite{beneventano,khedif2}. 
On the other hand, (\ref{ahcon}) can be interpreted as follows.
The number $4$ is accounting the degeneracy of spin and valley where
each one is contributing by $2$. 
${1\over 2}$ is coming from the Berry phase contribution that is equal $\pi$. This
makes difference with respect to the conventional QHE.

An important peculiarity of the Landau levels for massless Dirac fermions
is the existence of zero-energy states, i.e. $m=0$ in~(\ref{dspec}). Clearly,
this is unlike the usual 2DEG with parabolic bands where
the first Landau level is shifted by ${B}$. The existence of $m=0$ in~(\ref{dspec})
leads to an anomalous QHE instead of the conventional IQHE. This anomaly
is provided by the famous Atiyah--Singer index theorem~\cite{kaku}. A connection
between this theorem for the Dirac operator on a compact coset space
and higher dimensional QHE can be found in~\cite{brian}.

Returning now to discuss FQHE in graphene. In doing so,
Let us consider $M$-particles in LLL, which of course means that all $m_i=0$
with $i=1,\cdots,M$ and each $m_i$ corresponds to 
the spectrum~(\ref{dwafu}--\ref{dspec}).  The total wavefunction
of zero-energy Landau level~(\ref{dwafu0})
can be written in terms of the Slater determinant. This is
\begin{equation}\lb{nwps}
\psi(z,\bar{z})= \epsilon^{i_1 \cdots i_M} z_{i_1}^{n_1} \cdots z_{i_M}^{n_M}
\exp\left(-{B\over 4}\sum_i^M|z_i|^2\right)
\end{equation}
where $\epsilon^{i_1 \cdots i_M}$ is the fully
antisymmetric tensor and ${n_i}$ are integers. It is relevant to write this
wavefunction as Vandermonde determinant. We have  
\begin{equation}\lb{nwp}
\psi(z,\bar{z})= {\sf{const}}\,  \prod_{i<j}^M\left(z_i-z_j\right)
\exp\left(-{B\over 4}\sum_i^M|z_i|^2\right).
\end{equation}
This can be interpreted by remembering the Laughlin wavefunction 
\begin{equation}\lb{lw}
\psi_{\sf Laugh}^l(z,\bar{z})=  \prod_{i<j}^M
\left(z_i-z_j\right)^{2l+1}\exp\left(-{B\over 4}\sum_i^M|z_i|^2\right).
\end{equation}
It is well-known that it has many interesting features and good ansatz to describe
the fractional QHE at   the filling factor $\nu ={1\over 2l+1}$,
with $l$ is integer value. It is clear that~(\ref{nwp}) 
is nothing but the first Laughlin state that corresponds to $\nu=1$. Actually,~(\ref{nwp})  
is describing the first quantized Hall plateau of the integer QHE. 
Note that~(\ref{lw}) can also be written as 
\begin{equation}\lb{lw2}
|l\rangle= \left\{\epsilon^{i_1 \cdots i_N} z_{i_1}^{n_1} \cdots z_{i_N}^{n_N}\right\}^{2l+1}
|0\rangle.
\end{equation} 
Consequently, the wavefunction for the particles in the graphene LLL 
are identical to those of the 2DEG LLL. We conclude that
the basic physics in both systems is the same. More
discussion about this issue can be found in~\cite{toke}.

For the higher Landau levels, different attends 
are proposed in dealing with the anomalous QHE in
graphene. Among them, we cite the $SU(4)$ wavefunction
realized by Boerbig and Regnault~\cite{goerbig2} as a candidate 
to describe the effect. In doing this, they extended
the Halperin wavefunction~\cite{halperin} to take account of
the spin and valley degree of freedom. This allowed them
to discuss different issues and predict some
filling factors. This wavefunction is related to that we
have proposed~\cite{schreiber}, in general way, by
using the matrix model
and non-commutative Chern--Simons theories.
In section 4, we give different discussions
about the matter.

\section{Sphere analysis}

After describing the anomalous QHE on plane $\mathbb{R}^2$,
we now consider the case of two-sphere $\mathbb{S}^2$. In fact,
we are attending to
construct the same effect and therefore generalize
all obtained results by considering graphene
as a spherical manifold. In doing this task,
we should first establish some mathematical tools.
This will allow us to write the Pauli--Schr\"odinger
Hamiltonian in terms of that of Landau. Using the same
technical as for  $\mathbb{R}^2$,
we derive the Dirac eigenvalues as well as its eigenfunctions.
These materials will be the subject of the present section.

\subsection{Pauli--Shr\"odinger Hamiltonian on sphere} 

To generalize the Pauli--Shr\"odinger Hamiltonian from $\mathbb{R}^2$ to
$\mathbb{S}^{2}$ of radius unit, we first start 
by defining some  tools on this compact surface. Indeed, the $\mathbb{S}^{2}$
 Kh\"ahlerian metric and the corresponding volume measure are 
\beq 
ds^2_{\sf s} =
\frac{4}{\left(1+ {z\cdot \bar z }\right)^2}dz \otimes d\bar z, \qquad
d\mu_{\sf s}(z) =
\frac{4d\mu_{\sf p}(z)}{\left(1+ {z\cdot \bar z }\right)^2}.
\label{metric}
\eeq
They go to the standard distance and the Lebesgue measure 
$4d\mu_{\sf p}(z)=dz d\bar z$ 
on the planar limit. With (\ref{metric}), the inner product
on $\mathbb{S}^{2}$ of two functions
$\Psi_1$ and
 $\Psi_2$ in the Hilbert space reads as
 \begin{equation} \lb{ipb1}
\langle\Psi_1 |\Psi_2\rangle =\int_{\mathbb{S}^2}
d\mu_{\sf s}(z) \bar{\Psi}_1 \times \Psi_2.
\end{equation}
Because of its relation to magnetic field, one has
to define the vector potential on $\mathbb{S}^{2}$.
This can be done by introducing
the differential one-form $\theta_{\sf s}$, such as
\beq
\theta_{\sf s}(z) = \frac{i(\bar zdz -zd\bar
z)}{1+ {z\cdot \bar z}}.
\label{kahlerform}
\eeq 
Combining this with the Dirac quantization
\beq
\int_{\mathbb{S}^{2}} F =\int_{\mathbb{S}^{2}} dA =2\pi k 
\eeq
we end up with the required vector potential 
\beq
A= {k\over 2} \frac{i(\bar zdz -zd\bar
z)}{1+ z\cdot \bar z }\equiv {k\over 2}\theta_{\sf s}(z).
\eeq
Its complex components are given by
\beq\lb{vpcom}
A_z = i  {k\over 2}\frac{\bar z} {1+ {z\cdot \bar z }}, \qquad
A_{\bar z} = -i{k\over 2} \frac{z} {1+ {z\cdot \bar z}}.
\eeq 
This is showing an interesting relation 
between the integer $k$ and magnetic field $B$, 
which will be used in discussing QHE. It is 
\beq
k= 2B.
\eeq
This also express the Landau level degeneracies of $N$-particles.
Moreover, it can be used to define the filling factor, such as 
\beq
\nu = {N\over k}.
\eeq

There are two ways to 
write down the Pauli--Schr\"odinger Hamiltonian $H_{\sf s}^{\sf PS}$
on $\mathbb{S}^{2}$. Indeed algebraically, it
can be done by introducing the suitable
covariant derivatives in terms of $\theta_{\sf s}$. In analogy to
 $\mathbb{R}^{2}$, we realize them as
\beq\lb{codi}
D_{\sf s}= \pa + i{k\over 2} \left(\mbox{ext}\theta_{\sf s}\right), \qquad
D_{\sf s}^{*} = \pa^{*}  - i  {k\over 2}\left(\mbox{ext}\theta_{\sf s}\right)^{*} 
\eeq
which act on the $p$-forms of $\mathbb{S}^{2}$. Therefore, the required
Hamiltonian can be mapped as
\beq
H_{\sf s}^{\sf PS}= D_{\sf s}^{*} D_{\sf s} + D_{\sf s} D_{\sf s}^{*}. 
\label{GPH}
\eeq 
This mapping has two advantages. Indeed, firstly it
can be used to make a group theory analysis of the present
work in similar way to those have been developed in~\cite{karabali,daoud}.
Secondly, it allows us to apply the spectral theory approach~\cite{jellal}
in order to deal with different issues. This requires to introduce a  
$ H_{\sf s}^{\sf PS}$ form in terms of the local coordinates $(z, \bar z)$
instead of operators~(\ref{codi}). To obtain such form, it is convenient to
consider the corresponding smooth differential one-forms as two component spinors 
\beq\lb{1forms}
\Phi = 
\left(\begin{array}{c}
 \Phi_1 \\ \Phi_2
\end{array} \right)
\eeq
which define a Hilbert space of the form
\beq
{\mathcal H}_{\sf s}=L^2\left(\mathbb{S}^2, d\mu_{\sf p}\right)dz \oplus
L^2\left(\mathbb{S}^2, d\mu_{\sf p}\right)d\bar z.
\eeq
By acting (\ref{GPH}) on (\ref{1forms}), we find an elliptic
operator 
\beq 
H_{\sf s}^{\sf PS}=
\left (\begin{array} {cc} H^{B ,B_1}_{\sf s} & 0\\
0 & H^{B_2,B}_{\sf s}
\end{array}\right )- {k\over 2}
\left (\begin{array} {c c} 1 & 0
\\ 0 & -1 \end{array}\right ).
  \label{explicitPH}
\eeq
The operator 
$H^{\al,\be}_{\sf s}$ is related to the Landau Hamiltonian
on $\mathbb{S}^2$ and they coincide under some conditions as
we will see next. 
It is given
by
\beq\lb{glham} 
H^{\al,\be}_{\sf s}=
\left(1+{z\cdot \bar z }\right) 
\left\{-\left(1+{z\cdot \bar z } \right)
\frac {\partial ^2}{\partial z\partial {\bar z}} - \al z\frac {\partial
}{\partial z}+ \be
{\bar z}\frac{\partial} {\partial {\bar z}}\right\}
+\alpha\beta z\cdot \bar z.
\eeq
where the fields $B_i$ reads as
\beq\lb{tfiel}
B_1= B-{2}, \qquad B_2= {B}+2 .
\eeq
We end this part by citing three remarks. Firstly,
 the mapping (\ref{explicitPH}) is helpful is sense that we can
derive the basic features of $ H_{\sf s}^{\sf PS}$ from those
of the Landau Hamiltonian on $\mathbb{S}^2$. Secondly, 
the $ H_{\sf s}^{\sf PS}$ form
is consistent in describing particles on a spherical graphene. 
In fact, recalling that graphene is resulting from an unification 
of two-subsystems. They form  
an honeycomb and each part is governed by
a  component of $ H_{\sf s}^{\sf PS}$. Globally, the system can be seen
as composite fermions, we will return to discuss these matters.

\subsection{Spectrum of the Hamiltonian $H^{\sf L}_{\sf s}$}

Staring from (\ref{explicitPH}), it is clear that we first need
to determine the Landau spectrum on  $\mathbb{S}^2$. To do this,
we introduce the corresponding Landau Hamiltonian. It
can simply be obtained by 
requiring $\al=\be\equiv {k\over 2}$ in (\ref{glham}), such as
\beq\lb{lhams} 
H^{\sf L}_{\sf s} = \left(1+ { z\cdot \bar z }\right) 
\left\{-\left(1+ {z\cdot \bar z }\right)\frac {\partial^2}{\partial
z\partial {\bar z}} - {k\over 2} \left(z\frac {\partial
}{\partial z} -
{\bar z}\frac{\partial} {\partial {\bar z}}\right)\right\} + 
{k^2\over 4} z\cdot \bar z.
\eeq 
Firstly, (\ref{lhams}) has been used by Haldane~\cite{haldane} 
to build a theory for the filling factor $\nu={1\over 2l+1}$ on $\mathbb{S}^2$
that generalizes the Laughlin one~\cite{laughlin} one on $\mathbb{R}^2$. 
Subsequently, it has been
generalized to the higher dimensional spaces by Karabali and Nair~\cite{karabali}
as well as the Bergman ball~\cite{daoud,jellal}.

Solving the eigenvalue problem  
\beq
H^{\sf L}_{\sf s}  \Psi =E \Psi 
\label{EVL} 
\eeq
one can show that the eigenvalues are given by
\beq 
E_{m}^{\sf s}= {k\over 2} (2m+1)
+  m(m+1), \qquad 0\leq m \leq k+1.
\label{LSpec}
\eeq 
To derive the corresponding eigenfunctions, 
we express $D_{\sf s}$ and $D_{\sf s}^{*}$
in terms of the local coordinates $(z,\bar z)$.
Thus from (\ref{vpcom}) and (\ref{codi}),
one can write 
\beq\lb{cdlco}
D_{\sf s} = (1+ {z\cdot \bar z }) {\pa \over \pa{\bar z}} - {k\over 2}z,
\qquad
D_{\sf s}^* = -(1+ {z\cdot \bar z }) {\pa \over \pa{z}} - {k\over 2}\bar z
\eeq
with the condition
\beq\lb{conds2}
\left[ D_{\sf s},  D_{\sf s}^*\right]= k.
\eeq
As usual,  we start by  
determining the fundamental state. This can be obtained by
integrating the lowest Landau condition
\beq
D_{\sf s} \Psi= 0.
\eeq
After calculation, we find 
\beq\lb{fonst}
\Psi(z,\bar z) = \left(1+{z\cdot \bar z} \right)^{-{k\over 2}} h(z)
\eeq
where $h(z)$ is holomorphic function of $z$ and must be a 
polynomial of order $\leq k$. Therefore, other eigenfunctions
can simply be obtained by successive application of the
raising operators on (\ref{fonst}). Doing this process,
we obtain
\beq\lb{oeifun}
\Psi_m(z,\bar z) = (D_{\sf s}^*)_{k-m} 
(D_{\sf s}^*)_{k-m+1} \cdots  (D_{\sf s}^*)_{k-1} 
\left[\left(1+{z\cdot \bar z} \right)^{-{k\over 2}} h(z)\right].
\eeq
These define a Hilbert space of dimension $k+2m+1$.
Note that, (\ref{oeifun}) can explicitly be written in terms of the
coordinates $(z,\bar z)$.

Since the above results generalize those of the Landau problem
on the flat surface, it is relevant to check the asymptotic behavior.
This can be achieved by requiring the planar limit. In this case,
we recover the usual energy levels
on plane as well as the corresponding eigenfunctions seen before.

\subsection{Spectrum of the Hamiltonian $H^{\sf PS}_{\sf s}$}

The above results will be employed to derive the spectrum of
$H^{\sf PS}_{\sf s}$. For this, we can split
the mother eigenvalue equation
\beq
 H^{\sf PS}_{\sf s}\Psi =E \Psi 
\label{EVP} 
\eeq
into two daughter's where the functions
$\Phi$~(\ref{1forms}) form a Hilbert space ${\mathcal H}_{\sf s}$,
 $\Phi_1$ and $\Phi_2$ are in the space $L^2(\mathbb{S}^2, d\mu_{\sf p})$. 
To clarify our statement, we can introduce an appropriate
unitary transformation. It is generated  by 
a $2\times 2$ matrix
\beq\lb{utran}
U
= \left( \begin{array} {c c}
1+{z\cdot \bar z}  & 0 \\ 0 & 1+ { z\cdot \bar z}
 \end{array}\right ).
\eeq
Now we can easily establish an explicit relation between $H^{\sf PS}_{\sf s}$
and the Landau Hamiltonian $H^{\sf L}_{\sf s}$. Therefore, we show that
$H^{\sf PS}_{\sf s}$ can be written as
\beq 
H^{\sf PS}_{\sf s} = U^{-1}\set{
 \left ( \begin{array} {c c}
H^{{\sf L},B_1}_{\sf s}
 & 0 \\ 0 &
H^{{\sf L},B_2}_{\sf s}
 \end{array}\right )
 -  \left (\begin{array} {c c}
{k\over 2} -1 & 0
\\ 0 & -\left({k\over 2} +1\right)  \end{array}\right )} U.
\label{UPH}
\eeq
Clearly, (\ref{utran}) tells us that any function $\Phi_i\in L^2(\mathbb{S}^2, d\mu_{\sf p})$ 
can be written in terms of another one belongs to the space 
$L^2(\mathbb{S}^2, d\mu_{\sf s})$. This is
\beq
\tilde{\Phi}_i = \left(1+ z\cdot \bar z\right)\Phi_i.
\eeq
Starting from (\ref{utran}), one can see that $H^{\al,\be}_{\sf s}$
transforms as 
\beq
H^{\al,\be}_{\sf s}\Psi = \left(1+ z\cdot \bar z\right)^{-1} 
H^{{\sf L}, {\al+\be\over 2}}_{\sf s} 
\left[\left(1+ z\cdot \bar z\right)\Psi\right]
\label{reductionL}
\eeq
where $H^{{\sf L}, {\al+\be\over 2}}_{\sf s} $ is nothing but
the Landau Hamiltonian on $\mathbb{S}^2$ with a magnetic field 
of the form ${\al+\be\over 2}$, which is equal to $B$.

With the above tools, we are able to determine
the spectrum of $H^{\sf PS}_{\sf s}$. Indeed, 
instead of (\ref{EVP}) we have two eigenvalue equations
given by
\beq 
H^{{\sf L},B_1}_{\sf s} \tilde{\Psi}_1=(E+4B _1) \tilde{\Psi}_1,
\qquad 
H^{{\sf L}, B_2}_{\sf s  }\tilde{\Psi}_2= (E -4B_2)\tilde{\Psi}_2
\label{SystEVP}
\eeq
where  $\tilde{\Psi}_1$ and $\tilde {\Psi}_2$ are in $L^2(\mathbb{S}^2,d\mu_{\sf s})$.
It is clear that the whole eigenvalue problem becomes that of Landau Hamiltonian
and then the required spectrum can easily be deduced.
Indeed, 
after some calculation, we show that the eigenvalues 
of  $H^{{\sf L},B_1}_{\sf s}$ are 
\beq 
E^+_{m}= B  m +  m(m-1),
\qquad 0 \leq m\leq k_1 +1. 
 \label{EV+}
\eeq 
In contrast, for the second magnetic field, we find 
\beq
E^-_{m'}= B (m'+1) +  (m'+1)(m'+2),
\qquad 0 \leq m'\leq k_2 +1.
\label{EV-}
\eeq 
To obtain the  $H^{\sf PS}_{\sf s}$ spinors, we start
by evaluating theirs components. This can be done by 
considering the lowering and raising operators
in terms of $k_i$ fields. They are
\beq\lb{cdlco}
D_{\sf s}^{k_i} = (1+ {z\cdot \bar z }) {\pa \over \pa{\bar z}} - {k_i\over 2}z,
\qquad
(D_{\sf s}^{k_i})^* = -(1+ {z\cdot \bar z }) {\pa \over \pa{z}} - {k_i\over 2}\bar z.
\eeq
Thus, the ground-state is solution of the equation
\beq
D_{k_i} \chi=0.
\eeq
This yields to the function
\beq
\chi (z,\bar z)=  \left(1+z\cdot \bar z\right)^{-{k_i\over 2}} h(z)
\eeq
where $h(z)$ is an holomorphic function of degree
$\leq k_i$ on $\mathbb{S}^{2}$.
As far as  (\ref{SystEVP}) is concerned, the eigenfunctions
can be obtained by a successive application of the 
corresponding raising operators. Indeed, for the first
spectrum, we have
\beq\lb{lpsi}
\Psi_{m} (z,\bar z) = \left(1+z\cdot \bar z\right)^{-1} 
(D_{\sf s}^{k_1})_{{k\over 2}-1}^*
(D_{\sf s}^{k_1})_{{k\over 2}}^*\cdots
(D_{\sf s}^{k_1})_{{k\over 2} +m-2}^*
\left[\left(1+z\cdot \bar z\right)^{-{k\over 2}-m+1}h(z)\right].
\eeq
Similarly, we obtain for the second Hamiltonian
\beq\lb{lppsi}
\Psi_{m'} (z,\bar z)=
\left(1+z\cdot \bar z\right)^{-1} 
(D_{\sf s}^{k_2})_{{k\over 2}+1}^*
(D_{\sf s}^{k_2})_{{k\over 2}+2}^*
\cdots
(D_{\sf s}^{k_2})_{{k\over 2}+m'}^*
\left[\left(1+ z\cdot \bar z\right)^{-{k\over 2}-m'-1}h(z)\right].
\eeq
Finally, the Hilbert space of the Pauli--Schr\"odinger 
Hamiltonian reads as
\beq
{\cal H}^{\sf PS}_{\sf s}=\set{\Psi_{m,m'} =
\left
(\begin{array} {c} \Psi_m\\ \Psi_{m'}
 \end{array}\right )\in L^2(\mathbb{S}^2,d\mu_{\sf p}) \oplus 
L^2(\mathbb{S}^2,d\mu_{\sf p}),
 \quad H^{\sf PS}_{\sf s} \Psi_{m,m'} =
E^{\pm}_{m,m'} \Psi_{m,m'}}.
\eeq
Note that, with these results one can also discuss 
the conventional QHE and see the effect of adding the spin
as a degree of freedom. This can be done,
for instance, by evaluating the density of particles
 and two-point function, but this discussion
is out of the scope of the
present paper.

\subsection{Dirac spectrum on two-sphere}

Motivated by the fact that the Dirac Hamiltonian $H_{\sf s}^{\sf D}$ is
the basic tool that can be used to discuss the anomalous QHE, we have
to build this later on $\mathbb{S}^2$. This can be realized
by applying the same technical as it has been done
for the $\mathbb{R}^2$ case. More precisely, 
we suggest to write the $H_{\sf s}^{\sf D}$ in terms
of differential operators those are related to
the Pauli--Schr\"odinger formalism
on $\mathbb{S}^2$. This suggestion allows us to
make an easily derivation of the $H_{\sf s}^{\sf D}$
eigenvalues and the corresponding eigenfunctions.

The Dirac operator on $\mathbb{S}^2$ can be written in different
forms~\cite{stern}. 
For our purpose, it is convenient to introduce the Hamiltonian
for one-pseudospin component as 
\begin{equation}\lb{diracs}
H_{\sf s}^{\sf D}= i
\left(\begin{array}{cc}
0 & D_{\sf s}\\
D_{\sf s}^* & 0
\end{array}\right)
\end{equation}
where lowering $D_{\sf s}$ and the raising  $D_{\sf s}^*$ operators 
are those given in~(\ref{cdlco}), which verify the commutation
relations similar to~(\ref{conds2}). 
Clearly, $H_{\sf s}^{\sf D}$ coincides exactly with its partner in
the planar limit. 
Now it is easy to see that the square of (\ref{diracs})
\begin{equation}\lb{diracss}
\left(H_{\sf s}^{\sf D}\right)^2= 
\left(\begin{array}{cc}
 D_{\sf s} D_{\sf s}^*& 0\\
0 & D_{\sf s}^* D_{\sf s}
\end{array}\right)
\end{equation}
is related the Pauli--Schr\"odinger Hamiltonian on $\mathbb{S}^2$. 
More precisely, $H_{\sf s}^{\sf D}$ is expressed in  terms of the
Landau Hamiltonian $H_{\sf s}^{\sf L}$ on $\mathbb{S}^2$.
This tells us that
the spectrum of the (\ref{diracs}) can be deduced from that of 
$H_{\sf s}^{\sf L}$. 

Consequently, using the general statement shown before one
can find the eigenvalues
\beq 
E_{m}^{\sf D}= \pm  \sqrt{|m (k+m)|}
 \label{EV+}
\eeq 
with the condition
$m\geq 1$ when $k=0$
The corresponding eigenfunctions are
\beq\lb{dsawf}
\Psi_{m}=
\left(\begin{array}{c}
-{\rm sgn} (m)i\psi_{|m|-1} \\
  \psi_{|m|}
\end{array}\right)
\eeq
where the wavefunctions $\psi_{|m|}$ are those given in (\ref{oeifun}).
In addition, there is a zero-energy mode whose eigenfunction
is given by
\beq\lb{ds0wf}
\Psi_{0}=
\left(\begin{array}{c}
0 \\
  \psi_{0}
\end{array}\right).
\eeq
This corresponds to the graphene LLL. In fact,
these results will applied to discuss the anomalous FQHE. More precisely,
we analysis the incompressibility in LLL as well as the particle density.
Furthermore, we show that how the obtained eigenfunction can be used to
construct a $SU(N)$ wavefunction and its links to other theories. 
All these materials will be
the subject of two last sections.

\section{Anomalous QHE on sphere}

We discuss the anomalous QHE on two-sphere $\mathbb{S}^2$. For this, 
we focus on two aspects of QHE, which concern
the composite fermion picture and the incompressibility.
For the first one, we show that the Pauli--Schr\"odinger system
has a composite fermion nature. For the second one, we
end up with a result that has been showed by
Karabali and Nair~\cite{karabali} in analyzing 
the conventional QHE on $\mathbb{S}^2$.

\subsection{Composite fermion picture}

To give an explanation of the filling factors beyond the
Laughlin states, Jain has introduced the composite fermion
formalism~\cite{cf}. In fact, they
 are new kind of particles appeared in
condensed
matter physics to provide an explanation of the behavior of particles
moving in the plane when a strong magnetic field $B$
 is present. Particles
possessing $2p$, with $p\in N^*$, flux quanta (vortices) can be
thought of being composite fermions. One of the most important features of
them is they feel effectively a magnetic field of the form
\beq
\lb{cfm}
B^*=B \pm 2p\Phi_0\rho
\eeq
where in our convention the unit flux is $\Phi_0= 2\pi$.
Recalling the relation between the filling factor $\nu$ and $B$,
one can define a similar quantity for the field $B^*$. This can be written as
\beq\lb{defi}
\nu^* = 2\pi {\rho\over B^*}.
\eeq
It is clear that from (\ref{cfm}), the factors $\nu$ and
$\nu^*$ can be linked to each other through
\beq\lb{rnns}
\nu ={\nu^*\over 2p\nu^* \pm 1}.
\eeq
This relation has been used to deal with different issues
related to QHE in 2DEG. More discussions about the mapping
 (\ref{rnns}) and its applications can be found
in~\cite{cf}.

As an immediate consequence of (\ref{rnns}), we can map
the anomalous FQHE in terms of IQHE in graphene. Indeed, 
it is easy to see
\beq\lb{nunus}
\nu^{\sf G}= 2{2n+1\over 4p(2n+1) \pm 1}.
\eeq
This result has been obtained in different contexts, 
for instance see~\cite{castroneto}. It is obvious that for
$n=0$, we obtain Laughlin states and the anomalous IQHE
can be recovered by fixing $p=0$. Moreover, (\ref{nunus})
tells us that the Jain's series is quite different from the
2DEG. Therefore, it is interesting to focus on FQHE in graphene.

The above results can be linked to the present work. Otherwise,
the composite fermion picture exists in our analysis. Indeed, if one 
look at the obtained fields in rearranging the Pauli--Schr\"odinger
Hamiltonian, one can notice that both fields $B^{\pm}$  behave as
effective fields in similar way to the composite fermion
field $B^*$ given in (\ref{cfm}). To clarify this 
statement, firstly recall that after applying an external field
 $B$, we finally got  $B^{\pm}$. Secondly, 
the Dirac quantization for these field can be deduced from
that of $B$. Thus, we have
\beq
k_1 = k- 4, \qquad k_2 =k+ 4.
\eeq
This is should be valid for any 
integer value $l$ because their summation
gives the whole external magnetic field
\beq
k_1 + k_2 =k.
\eeq
 Thirdly, let us write the fields as 
\beq\lb{obfd}
k^{\pm} = k \pm l
\eeq 
where we have set $-\equiv 1$ and $+\equiv 2$.
Now it becomes clear that  
there is a shift with respect to $B$, which is analogue to (\ref{cfm}).
Therefore, (\ref{obfd}) shows that 
the present system behaves as a collection of composite
fermions where $l$ is attributed to vortices. 
Moreover, we can define the filling factor $\nu$
in terms of the quantity $\nu^{\pm}$, such as
\beq\lb{rnunump}
\nu ={\nu^{\pm}\over 2p\nu^{\pm} \pm 1}.
\eeq
Adopting the definition (\ref{defi}), we can express  $\nu^{\pm}$
as 
\beq
\nu^{\pm} =2\pi {\rho\over B^{\pm}} \equiv 2\pi {\rho} l_{B^{\pm}}^2
\eeq
where $2\pi l_{B^{\pm}}^2$ is the Hall droplet area of the
composite fermions submitted to $B^{\pm}$. 
At this level, one can give 
some interpretations to (\ref{rnunump}) by switching on 
 $\nu^{\pm}$ to different values. Indeed,
for $\nu^{\pm} =1$ we recover Laughlin states~\cite{laughlin}
and  for $\nu^{\pm} $ integer we end up with Jain sequences~\cite{cf}.

\subsection{Graphene LLL}

Particles in  LLL are confined in a potential
that is grand enough to neglect the kinetic energy. This
produces a gap such that  particles are not allowed to jump to
the next level. LLL is rich and contains many interesting
features that are relevant in discussing QHE. 
With this, it is interesting to
consider the present system on LLL. 
Before talking about the graphene LLL for two-sphere,
let us make some discussions about LLL of the Landau and Pauli--Schr\"odinger
Hamiltonian's.

We keep particles in LLL and investigate their basic features. 
We start by giving the
 spectrum for one particle in LLL, which can be obtained 
just by fixing $m=0$ in the previous analysis. This gives 
the ground state in (\ref{fonst})
that has the energy
\begin{equation}\lb{0eb1}
E_0= {B}.
\end{equation}
This  value coincides with  that corresponds to  the
Landau problem on the plane. More discussions about the 
issue can be found in Karabali and Nair work~\cite{karabali}.

One particle spectrum in LLL can be generalized to
that for $M$-particles. It is obvious that the total energy is given by 
\begin{equation}\lb{N0eb1}
 E_M= {MB}.
\end{equation}
The corresponding wavefunction can be constructed as the Slater determinant. This is 
\begin{equation}\lb{Ngsb1}
\Psi_M(z)= \epsilon^{i_1 \cdots i_M}
\Psi_{i_1}(z_{i_1})  \Psi_{i_2}(z_{i_2})
\cdots  \Psi_{M_1}(z_{i_M})
\end{equation}
where each  $\Psi_{i_j}(z_{i_j})$ has the form given
in~(\ref{fonst}). This is similar to the wavefunction
(\ref{nwps}) on the plane and  corresponds to the filling 
factor $\nu=1$. Other Laughlin states can be obtained as 
we have given in~(\ref{lw2}). 

 Now let us turn to the Pauli--Schr\"odinger Hamiltonian $H_{\sf s}^{\sf PS}$. First,
we emphasis an important behavior of its spectrum. Indeed,
unlike the Landau Hamiltonian, $H_{\sf s}^{\sf PS}$ has an isolated
eigenvalue and in this case, the Hilbert space becomes
\beq
{\cal H}^{\sf D}_{\sf s}=\left\{
\left
(\begin{array} {c} \Phi_0\\ 0
 \end{array}\right ), \quad 
\Phi_0= (1+z\cdot \bar z)^{-{k\over 2}} h(z)\right\}.
\eeq
It is describing the first component LLL of $H_{\sf s}^{\sf PS}$.
In contrast, the second LLL is governed by the energy
\beq
E^-_{m'} = {B} +2 \equiv B_2
\eeq
which is not equal to zero. However this is nothing but
that corresponding to plane results for a composite 
fermion subjected to the magnetic field $B_2$.
Therefore, all discussions reported on QHE in $\mathbb{S}^2$
can be applied to the second component of $H_{\sf s}^{\sf PS}$.

At this point, we return  to analysis the anomalous QHE on $\mathbb{S}^2$.
As we noticed before, the graphene LLL is described by a
zero-energy mode that corresponds to the wavefunction~(\ref{fonst}).
Since this is also the same state for the Landau Hamiltonian on 
$\mathbb{S}^2$ then we can construct the same Langhlin states~(\ref{Ngsb1})
for the filling factor ${1\over 2l +1}$.
Therefore, we have the same results as those obtained
in the conventional QHE on $\mathbb{S}^2$. In fact, we can show that
the density of particles is constant and the the probability of finding 
two particles at the same position is zero. These basically the results
obtained by Karabali and Nair~\cite{karabali} in analyzing the
conventional QHE on the projective complexes spaces $\mathbb{CP}^d$,
with $d$ is an integer.

We now underline that there is no difference between the graphene LLL
and the 2DEG LLL for a compact surface. Indeed,
the definition (\ref{defi}) tells us that the density of particles is an important
ingredient. To get QHE, this parameter should be kept constant
by varying  the magnetic field. For its relevance, we evaluate the
density, which is
\begin{equation}
\rho ={M\over 4\pi r^2} = {BM\over 4 \pi k}
\end{equation}
for a two-sphere of volume $4\pi r^2$.
In the thermodynamic limit, i.e. $(M, k \lga \infty)$, it goes to 
the finite quantity 
\beq
\rho \sim {B\over 2\pi}. 
\eeq
This is exactly the density of particles on flat geometry and therefore
 corresponds to the fully occupied state 
 $\nu=1$.

In QHE,  the quantized plateaus come from the realization of an incompressible liquid.
 This property is important since it is related to the energy. It 
means that by applying an infinitesimal 
pressure to an incompressible system the volume remains unchanged~\cite{ezawa}. 
This condition can be checked for our system by
considering two-point function and integrating over
all particles except two.  This is 
\begin{equation}\lb{tpfb1}
I(z_{i_1},z_{i_2})= \int_{\mathbb{S}^2} 
d\mu_{\sf s}\left(z_3,z_4,\cdots, z_M\right)
\left[\Psi_M(z)\right]^{*} \Psi_M(z).
\end{equation}
It is easy to see that $I(z_{i_1},z_{i_2})$ reads as
\begin{equation}\lb{2tpfb1}
I(z_{i_1},z_{i_2})\sim |\Psi_{0,i_1}|^2 |\Psi_{0,i_2}|^2-
|\left(\Psi_{0,i_1}\right)^*\Psi_{0,i_2} |^2.
\end{equation}
This can also be evaluated in the planar limit as 
\beq
\lb{22tpfb1}
I(z_{i_1},z_{i_2})\sim 1- \exp\left[- k|\vec x_1 -\vec x_2|^2 \right]
\eeq
which tells us that the probability of finding 
two particles at the same position is zero,
as it should be. This is analogue to the result obtained
by Karabali and Nair~\cite{karabali} for the conventional FQHE.

The above result showed that there is no difference between the
QHE at LLL in both systems: graphene and 2DEG. This suggests
to analyze the higher Landau levels and show if there is any
difference between them. To reply this inquiry, one may go straightforwardly
to generalize the Haldane wavefunction for the conventional
QHE to that for graphene.

\section{$SU(N)$ wavefunction}

Using different theoretical arguments, some authors suggested a 
possible FQHE in graphene. As we have seen before, this will be
natural if we look at the anomalous IQHE as a product of
 collective behavior of the composite fermions instead of
particles. Moreover, different wavefunctions have been proposed
to describe AFQH for different filling factors. Among them, we cite
that constructed by Goerbig and Regnault~\cite{goerbig2} to solve some
problems brought by other theories like for instance that
developed in~\cite{toke}. However, this wavefunction is
sharing many common features with our early proposal~\cite{schreiber}.
On the other hand, the  Goerbig and Regnault states are a direct extended
version of those of the Laughlin~\cite{laughlin} as well as Halperin~\cite{halperin}.
These states are also translationally invariant but they are not rotationally.
To overcome this problem, one may apply the Haldane~\cite{haldane} picture to 
deal with the anomalous FQHE at the filling factor~\cite{wen}
\begin{equation}
\label{biff}
\nu=q_iK_{ij}^{-1}q_j
\end{equation}
where $K_{ij}$ is an $N\times N$ matrix and $q_i$ is a vector.
In fact, this will include different fractions suggested recently
for FQHE in graphene and allows us to make contact
with different proposals.

\subsection{Generic case}

To construct a general wavefunction, we use the obtained
result so far. In fact, the starting point is to note
that from the Haldane wavefunction one can realize that of Halperin
 on $\mathbb{S}^2$. In doing so,
we consider two sectors labeled by $(m)$ and $(n)$. This is the case for 
instance in graphene where there are two subsystems forming a honeycomb.
Le us define $\psi^{(m,n)}$ as a tensor product
\beq
\psi^{(m,n)} = \psi^{(m)} \otimes \psi^{(n)}
\eeq
where each $\psi^{(m)}$ is given by~(\ref{fonst}). Assuming that
the condition between matrix elements $K_{ij}=K_{ji}$ is
fulfilled, a natural way to construct the required 
 wavefunction is
\begin{eqnarray}
\label{nvacuum}
| \Psi \rangle &=& \prod_{m=1}^{N} 
\left[ \epsilon ^{i_1\cdots i_{M_m}} \psi_{i_1}^{(m)} \psi_{i_2}^{(m)} 
\cdots \psi_{i_{M_{m}}}^{(m)}  
\right]^{K_{mm}-K_{mn}}  \ \prod_{n=1}^{N} 
\left[ \epsilon ^{j_1\cdots j_{M_n}} \psi_{i_1}^{(n)} \psi_{i_2}^{(n)} 
\cdots \psi_{i_{M_{n}}}^{(n)}  
\right]^{K_{nn}-K_{mn}} 
\nonumber\\
&& 
\prod_{m<n}^{N}
\left[ \epsilon ^{k_1 \cdots k_{M_m+M_n}} \psi_{k_1}^{(m,n)} \psi_{k_2}^{(m,n)} 
 \cdots \psi_{i_{M_{m}+M_{n}}}^{(m,n)}  
\right]^{K_{mn}}  | 0 \rangle.
\end{eqnarray}
We can show that $| \Psi \rangle$ verifies the constraint
\begin{equation}
L_{\sf tot}|\Phi \rangle =0
\end{equation}
where the total angular momenta is given by
\beq
L_{\sf tot}=\sum_{i}^N L_{i}.
\eeq
The components of each $L_{i}$ can be written in terms of the
global coordinates $u_i$ and $v_i$ as
\beq
L_i^+ = u_i{\pa\over \pa v_i}, \qquad L_i^- = v_i{\pa\over \pa u_i},
\qquad L_i^z = {1\over 2}\left(u_i{\pa\over \pa v_i} -v_i{\pa\over \pa u_i}\right)
\eeq
forming a closed Lie algebra of the $SU(2)$ group. 
On the other hand, novel about this vacuum configuration is that one can  
interpret the term 
\beq
\lb{INT}
\prod_{m<n}^{N}
\left[ \epsilon ^{k_1 \cdots k_{M_m+M_n}} \psi_{k_1}^{(m,n)} \psi_{k_2}^{(m,n)} 
 \cdots \psi_{i_{M_{m}+M_{n}}}^{(m,n)}  
\right]^{K_{mn}} 
\eeq
as an inter-layer correlation. In conclusion, our
configuration could be a good ansatz for 
the ground states of FQHE in graphene. This 
will be clarified soon.

To write the above wavefunction in terms of the global coordinates,
one may recall the Haldane realization~\cite{haldane} of the Laughlin
states~\cite{laughlin} on $\mathbb{S}^2$. This is
\beq\lb{hawf}
\Psi^l_{\sf H} = \prod_{i<j}^M \left(u_iv_j - u_jv_i \right)^{2l+1}
\eeq
where $u$ and $v$ are given by
\beq
u= \cos\left({\te\over 2}\right) \exp\left({i\over 2}\varphi\right),
\qquad 
v= \sin\left({\te\over 2}\right) \exp\left({i\over 2}\varphi\right).
\eeq
These are related to  the real coordinates $(x^1, x^2, x^3)$ on $\mathbb{S}^2$ 
through
\beq
x^i = \rho u^{\da} \si^i u
\eeq
where $\si^i$ are the Pauli matrices. 
Using the standard stereographic mapping,
we  express the global coordinates
in terms of $(z, \bar z)$ as
\beq\lb{uvzbz}
u_i = {1\over 1+ z_i\cdot \bar z_i}, \qquad v_i = {z_i\over 1+ z_i\cdot \bar z_i}.
\eeq

At this level, we have all ingredients to do our job. Indeed,
we start by defining
a new complex variable
\begin{equation}
\label{zeta}
\zeta_i=\left\{ 
\begin{array}{l}
z_{i}^{(m)} \qquad {\mbox{for}}\; i=1, \cdots, M\\
z_{{i-M}}^{(n)}\qquad {\mbox{for}}\; i=M+1,  \cdots,2M
\end{array}
\right.
\end{equation}
assuming that the particle numbers are 
equal, i.e. $M_1=M_2=M$, 
and recalling the antisymmetric Vandermonde determinant
for the fully occupied state
\begin{equation}
\prod_{i<j} \left(z_i-z_j\right) = {\rm det}\left(z_i^{M-j}\right)=
\epsilon^{i_1 \cdots i_M}z_{i_1}^{0}\cdots z_{i_M}^{M-1}.
\end{equation}
Therefore in the planar limit,~(\ref{nvacuum})
can be projected on the complex plane as
\begin{eqnarray}
\label{projection}
\Psi_{|_{\sf plane}} &=&
\prod_{m=1}^N  \left[ \epsilon ^{i_1\cdots i_M}  \left(z_{i_1}^{(m)}\right)^0 \cdots
 \left(z_{i_M}^{(m)}\right)^{M-1} \right]^{K_{11}-K_{12}}
\prod_{n=1}^N  \left[ \epsilon ^{j_1 \cdots j_M}  \left(z_{j_1}^{(n)}\right)^0 \cdots
\left(z_{j_M}^{(n)}\right)^{M-1} \right]^{K_{22}-K_{12}} 
\nonumber\\
&&
\prod_{m<n}^N \left[ \epsilon ^{k_1 \cdots k_{2M}}  \zeta_{k_1}^0 \cdots
\zeta_{k_{2M}}^{2M-1} \right]^{K_{12}}\; 
\Psi_{0}.
\end{eqnarray}
It can be written in the standard form as
\beq\lb{zcvacu}
\Psi_{|_{\sf plane}} =
\prod_{m=1}^N \prod_{i<j}^M \left(z_i^{(m)} - z_j^{(m)} \right)^{K_{mm}}
 \prod_{n=1}^N \prod_{i<j}^M \left(z_i^{(n)} - z_j^{(n)} \right)^{K_{nn}}
\prod_{m<n}^N \prod_{i,j}^M \left(z_i^{(m)} -z_j^{(n)} \right)^{K_{mn}} 
\; \Psi_{0}.
\eeq
This exactly coincides with that constructed by Goerbig and Regnault~\cite{goerbig2}
and what we have proposed~\cite{schreiber} in terms of the matrix model and
non-commutative Chern--Simom theories.
Using the mapping~(\ref{uvzbz}), it is easy to show that~(\ref{nvacuum})
takes the form
\begin{eqnarray}
\lb{cvacu}
\Psi_{\left(K_{mm},K_{nn},K_{mn}\right)} &=&
\prod_{m=1}^N \prod_{i<j}^M \left(u_i^{(m)} v_j^{(m)} - u_j^{(m)} v_i^{(m)}\right)^{K_{mm}}
 \prod_{n=1}^N \prod_{i<j}^M \left(u_i^{(n)} v_j^{(n)}- u_j^{(n)} v_i^{(n)}\right)^{K_{nn}}\\
\nonumber
&&
\prod_{m<n}^N \prod_{i,j}^M \left(u_i^{(m)} v_j^{(m)} -u_j^{(n)} v_i^{(n)}\right)^{K_{mn}} 
\; \Psi_{0}.
\end{eqnarray}
It clear that, the first and second part of~(\ref{cvacu}) are nothing but
two copies of Haldane wavefunctions for two systems
without interaction between each other. However, the third term
is showing the inter-layer correlation
\beq
\prod_{m<n}^N \prod_{i,j}^M \left(u_i^{(m)} v_j^{(m)} -u_j^{(n)} v_i^{(n)}\right)^{K_{mn}}.
\eeq
The wavefunction~(\ref{cvacu}) is a good candidate for describing the anomalous
FQHE in graphene. 
Next focusing on the $N=4$ case, we will illustrate~(\ref{cvacu}) 
by giving different applications. For this,
we give some configurations those lead to recover
some filling factors.

\subsection{$N=4$ case}

As stated before, the factor $4$  has different independent origins and together
contribute in the anomalous QHE. In fact, it can be regarded as
a manifestation of two spin states as well as two Dirac points.

{\underline{\bf Illustration 1:}}
Considering the two layers and treating them as
additional degrees of freedom, the $\nu={1\ov 2}$
state was predicted by Yoshioka, MacDonald and 
Girvin~\cite{girvin1}. They made
a straightforward generalization of the Laughlin
states to those with the filling factor
\begin{equation}
\lb{ymgf}
\nu={2\over k+l}
\end{equation}
where $k$ and $l$ are integers. This can be obtained   
from our generalization on $\mathbb{S}^2$ by taking
the configuration 
\begin{equation}
\begin{array}{l}
{K}=\left(\begin{array}{ll}
k & l\\
l & k\end{array}\right), \qquad
q=\left(\begin{array}{ll}
1&-1\end{array}\right)
\end{array}
\end{equation}
which is leading to the wavefunction 
\begin{eqnarray}
\lb{ycvacu}
\Psi_{(k,k,l)} &=&
\prod_{m=1}^4 \prod_{i<j}^M \left(u_i^{(m)} v_j^{(m)} - u_j^{(m)} v_i^{(m)}\right)^{k}
 \prod_{n=1}^4 \prod_{i<j}^M \left(u_i^{(n)} v_j^{(n)}- u_j^{(n)} v_i^{(n)}\right)^{k}\\
\nonumber
&&
\prod_{m<n}^4 \prod_{i,j}^M \left(u_i^{(m)} v_j^{(m)} -u_j^{(n)} v_i^{(n)}\right)^{l} 
\; \Psi_{0}.
\end{eqnarray}
It is clear that this is corresponding to different filling factors. Indeed,
choosing $k=3$ and $l=1$,
we recover the FQHE $\nu={1\over 2}$ state
corresponding to
\begin{eqnarray}
\lb{12cvacu}
\Psi_{(3,3,1)} &=&
\prod_{m=1}^4 \prod_{i<j}^M \left(u_i^{(m)} v_j^{(m)} - u_j^{(m)} v_i^{(m)}\right)^{3}
 \prod_{n=1}^4 \prod_{i<j}^M \left(u_i^{(n)} v_j^{(n)}- u_j^{(n)} v_i^{(n)}\right)^{3}\\
\nonumber
&&
\prod_{m<n}^4 \prod_{i,j}^M \left(u_i^{(m)} v_j^{(m)} -u_j^{(n)} v_i^{(n)}\right)^{1} 
\; \Psi_{0}.
\end{eqnarray}
Moreover, $\nu={2\over 3}$ can be recovered by setting 
for instance $k=2$ and $l=1$. Other filling factors can be 
derived in similar way.

{\underline{\bf Illustration 2:}}
By adopting the Halperin picture~\cite{halperin} for the conventional FQHE, 
another interesting result can be obtained. 
Indeed, in the 
context of single-layered unpolarized
QH systems, the labels $m$ and $n$
can be considered as an analogue of spin.
Following this idea, our graphene system can
be seen as mixing layers of particles with  
spin up and spin down.
As a consequence, for $k=3$ and $l=2$, we get  
the unpolarized Halperin 
wavefunction with the filling factor
${2\ov 5}$
\begin{eqnarray}
\lb{25cvacu}
\Psi_{(3,3,2)} &=&
\prod_{m=1}^4 \prod_{i<j}^M \left(u_i^{(m)} v_j^{(m)} - u_j^{(m)} v_i^{(m)}\right)^{3}
 \prod_{n=1}^4 \prod_{i<j}^M \left(u_i^{(n)} v_j^{(n)}- u_j^{(n)} v_i^{(n)}\right)^{3}\\
\nonumber
&&
\prod_{m<n}^4 \prod_{i,j}^M \left(u_i^{(m)} v_j^{(m)} -u_j^{(n)} v_i^{(n)}\right)^{2} 
\; \Psi_{0}.
\end{eqnarray}
This can be seen as the wavefunction of a system
of $M$-particles with spin parallel $m=\uparrow$
and $M$-particles with spin 
anti-parallel $n=\downarrow$ to the external magnetic field. 

{\underline{\bf Illustration 3:}} 
Finally, we give a
configuration that allows us to derive other interesting states like
${8\over 19}$. 
Indeed,
by requiring
\begin{equation}
\begin{array}{l}
{K}=\left(\begin{array}{ll}
k&l\\
l&k\end{array}\right), \qquad
q=\left(\begin{array}{ll}
2&-2\end{array}\right)
\end{array}
\end{equation}
we find another sequence of the filling factors
\beq
\nu={8\over m+n}.
\eeq
This is including
the state ${8\over 19}$ that may not be
described by some theories. It can be recovered
by fixing for instance $k=10$ and $l=9$.

\section{Conclusions}

We have analytically analyzed the Dirac Hamiltonian
with a magnetic field for particle living on two-sphere
$\mathbb{S}^2$. To do this, we have introduced the spectral
properties of the Pauli--Schr\"odinger Hamiltonian $H_{\sf s}^{\sf PS}$ and
explored its relationship to the Landau problem.
This has been done by making use of an appropriate transformation,
which is generated
by a $2\times 2$ matrix. The matrix elements of $H_{\sf s}^{\sf PS}$
have been seen as two Hamiltonian's describing  
two systems submitted to different portion of the external magnetic field.
Subsequently, after mapping the Dirac 
operator in terms of Landau Hamiltonian, it was easy to derive
its eigenvalues as well as the corresponding eigenfunctions.

We have shown that the Pauli--Schr\"odinger system is exhibiting
a composite fermion behavior. This has been done by establishing
a contact between our subfields $B^{\pm}$ and those felt by composite
fermions in the Jain picture~\cite{cf}. Moreover, using the
stand definition, we have written  a filling factor
in terms of $B^{\pm}$ and linked to different sequences.
In fact, we have reproduced the Laughlin states for $\nu^{\pm}=1$
and those of Jain for $\nu^{\pm}$ integer.

Before discussing the anomalous fractional quantum Hall effect
(AFQHE) on $\mathbb{S}^2$, we have made some notes on its partner
on plane. For this we have introduced some discussions related the
anomalous integer Hall conductivity and its extended version to
 AFQHE on $\mathbb{R}^2$. Subsequently, we have moved to
$\mathbb{S}^2$ in order to show what are the similarities and differences
between AFQHE in two-dimensional electron gas (2DEG) and graphene.
In the beginning we have focused
on the lowest Landau level (LLL), i.e. $m=0$ in Dirac spectrum, where two
physical quantities have been evaluated. These concerned the
incompressibility as well as the density of particles. These allowed us
to conclude that the basics physics in the graphene LLL is the same
as in the 2DEG LLL.

To analyze the high Landau levels, we have extended the Haldane wavefunction
on $\mathbb{S}^2$ to $SU(N)$ states those showing many interesting properties.
This has been done by using our early work~\cite{schreiber} related to the matrix model and
non-commutative Chern--Simons theories. Our states allowed us to generalize
their partners on the planar limit~\cite{goerbig2}
and describe different filling factors. Among them, we cite ${8\over 19}$ that
may not, in some instance, be described by the $SU(4)$ composite fermion
wavefunction~\cite{toke}.

The present investigation of the anomalous QHE on $\mathbb{S}^2$ can not
remain at this level. In fact, many questions can be extracted and solved
in different ways. Firstly, one may think to make a group theory
analysis of the present results in similar way to that have been reported by Karabali and
Nair~\cite{karabali} as well as our works~\cite{daoud}. Secondly, it might of 
interest to consider the anomalous effect on the three-sphere $\mathbb{S}^3$
by applying an approach used by Nair and Randjbar-Daemi~\cite{nair} to
underline what makes differences with respect to  the conventional QHE. 
Still also other questions where some of them are under preparation.

\section*{Acknowledgment} 

This work was completed during a visit of AJ to High Energy Section of
the Abdus Salam International Centre for Theoretical Physics
(AS-ICTP). He would like to thank Professor S. Randjbar-Daemi for
his kind invitation.  The author is indebted to the
referee for his constructive comment.

\end{document}